\documentclass[12pt,preprint]{aastex}

\slugcomment{Ap. J. (accepted, Jul. 8, 2004) }

\shorttitle{HC$_{3}$N ladders in CRL618}
\shortauthors{Pardo et al.}

\begin{document}


\title{The slowly expanding envelope of CRL618
probed with HC$_{3}$N rotational ladders}


\author{J. R. Pardo\altaffilmark{1}, J. Cernicharo\altaffilmark{1} and J.R. Goicoechea}
\affil{IEM - Departamento de Astrof\'{\i}sica Molecular 
              e Infrarroja, CSIC, Serrano 121, E-28006 Madrid, Spain.}

\and

\author{T.G. Phillips}
\affil{Division of Physics, Mathematics and Astronomy, California 
Institute of Technology, MS 320-47, Pasadena, CA 91125, USA}


\altaffiltext{1}{Visiting scientist at 
Division of Physics, Mathematics and Astronomy, California 
Institute of Technology, MS 320-47, Pasadena, CA 91125, USA}


\begin{abstract}
Lines from HC$_{3}$N and isotopic substituted species in 
ground and vibrationally excited states 
produce crowded millimeter and submillimeter wave 
spectra in the C-rich protoplanetary nebula CRL618. The complete 
sequence of HC$_{3}$N rotational lines 
from J=9--8 to J=30--29 has been observed with the IRAM 30m 
telescope toward this object. Lines from 
a total of 15 different vibrational states (including 
the fundamental), with energies up to 1100 cm$^{-1}$,  
have been detected for the main HC$_{3}$N isotopomer. In addition, the 
CSO telescope has been used to 
complement this study in the range J=31--30 to J=39--38, with detections 
in five of these states, all of them below 700 cm$^{-1}$. Only the rotational lines 
of HC$_{3}$N, in its ground vibrational state, display evidence of the well known 
CRL618 high velocity velocity outflow. Vibrationally excited 
HC$_{3}$N rotational lines exhibit P-Cygni profiles at 3 mm, 
evolving to pure emission lineshapes at shorter wavelengths. This evolution 
of the line profile shows little dependence on the vibrational state from which they 
rotational lines arise. The absorption 
features are formed against the continuum emission, which  
has been successfully characterized in this work due to the large frequency coverage. The  
fluxes range from 1.75 to 3.4 Jy in the frequency range 90 to 240 GHz.  
These values translate to an effective continuum source with a 
size between 0.22'' and 0.27'' and effective temperature at 200 GHz ranging from 
3900 to 6400 K with spectral index between -1.15 and -1.12. We have made an effort 
to simultaneously fit a representative set of observed HC$_{3}$N lines 
through a model for an expanding shell around the central star 
and its associated HII region assuming that LTE prevails for HC$_{3}$N. The simulations 
show that the inner slowly expanding envelope has expansion and turbulence 
velocities of $\sim$ 5-18 
kms$^{-1}$ and $\sim$ 3.5 kms$^{-1}$ respectively, and 
that it is possibly elongated. Its inclination with respect to the line of sight 
has also been explored. The HC$_{3}$N column 
density in front of the the continuum source has been determined by comparing 
the output of an array of models to the data. The best fits are obtained 
for column densities 
in the range 2.0-3.5$\cdot$10$^{17}$ cm$^{-2}$, consistent with previous estimates 
from ISO data,  and T$_{K}$ in the range 250 to 275 K, in very good agreement 
with estimates made from the same ISO data.

\end{abstract}




\section{Introduction}
CRL 618 is probably the best example of a C-rich protoplanetary nebula (PPNe) with 
a thick molecular envelope (Bujarrabal et al. 1988) surrounding a B0 star and an
ultracompact HII region from which UV radiation 
impinges on the envelope. Its distance of  
1.7 kpc makes this object one of the best suited sources 
for studying the evolutionary stages from the Asymptotic Giant Branch (AGB) 
to planetary nebulae (Herpin et al. 2002). The brightening of the HII region in 
the 1970s (Kwok \& Feldman 1981; Mart\'{\i}n-Pintado et al. 1993) and
the discovery of high velocity molecular winds (HVMW) with velocities 
up to 200 kms$^{-1}$ (Cernicharo et al. 1989) illustrate the rapid evolution 
of the central star and its influence on the circumstellar ejected material. 
From observations of molecular gas at arcsecs resolution, the high velocity 
molecular outflow and the slowly expanding envelope have been resolved  
and precisely located around the ultracompact HII region (Neri et al. 1992, 
Cox et al., 2003, S\'anchez-Contreras \& Sahai, 2004).

The recent discovery  of the polyynes HC$_{4}$H and HC$_{6}$H  
and of benzene (C$_6$H$_6$), the first aromatic molecule 
detected outside the solar system, in CRL618 
(Cernicharo et al., 2001a,b), emphasizes the fact that 
the copious mass ejection, toward the end of the AGB phase,  
and the related shock and UV driven chemistry, make the C-rich 
nebulae of this type very efficient factories of organic molecules 
(see detailed chemical models for this object in Cernicharo 2004). 
Small hydrocarbons and pure carbon chains are formed and they can in 
turn be the small nuclei from which larger C-rich 
molecules, such as the 
Poli-Aromatic Hydrocarbons (PAHs) can be built. These 
molecules are proposed as responsible for the observed emission 
labeled as Unidentified Infrared Bands (UIBs). 

In order to draw the clearest picture of the chemical composition, structure 
and dynamics of CRL618, it is necessary to perform complete line surveys in those 
regions of the electromagnetic spectrum where most molecules do emit. One of 
these line surveys has been carried out by us at millimeter wavelengths with 
the IRAM 30m telescope (Cernicharo et al., in prep.). One of the most important 
features seen in the survey is the presence of clusters of lines with a frequency 
separation of $\simeq$ 9 GHz, corresponding to rotational transitions in  
the ground and vibrationally excited states of HC$_3$N. This molecule belongs to 
the cyanopolyynes, HC$_{2n+1}$N, an important group of interstellar molecules 
(detected in space up to 2n+1=11, Bell et al. 1997). Because of the  
relatively low-energy of their bending modes, they can be found in space 
in many vibrationally excited states. As an example Wyrowski et al. (1999) 
detected HC$_{3}$N 
in eleven vibrationally excited states with energies up to 1100 cm$^{-1}$ toward 
the ultracompact HII region G10.47+0.03. The small frequency spacings of 
the HC$_{3}$N rotational transitions make this molecule an interesting tool to 
probe the physical conditions of molecular gas in C-rich circumstellar shells 
such as in CRL618. The HC$_{3}$N line profiles in this object show the interesting 
behavior of evolving, with frequency, from P-Cygni to normal emission. This 
is presented and analyzed in this paper using a model, under LTE, for an expanding 
envelope of azimuthally symmetric geometry, around a continuum source.

In section 2 we present the IRAM-30m and CSO observation procedures. The spectroscopy 
of HC$_{3}$N is briefly summarized in section 3. The results of the observations 
including discussions about the continuum emission and the velocity components of 
the HC$_{3}$N lines are found in section 4. In section 5 we present the 
model that has been built to explore the physical conditions. 
Simulations are presented and discussed. Finally, the conclusions are presented  
in section 6.

\section{Observations}
\label{sct:obs}
The CRL 618 observations presented in this paper are part of a line 
survey, carried out with the IRAM-30m telescope, and complemented with 
the Caltech Submillimeter Observatory (Cernicharo et al., in prep).

The IRAM-30m telescope has presently the largest effective aperture 
operating at the 3, 2 and 1.3 mm atmospheric windows.  
The observations started in 1994 and were completed in the winter of  
2001-02. In order to avoid 
confusion between the signal and the image sidebands, the receivers were 
optimized for image sideband rejection larger than 12 dB, which resulted   
in practically no sideband contamination, except 
for just a few cases where a very intense line (CO, HCN,...) was 
present in the image sideband. Receivers covered 
the following frequency ranges (in GHz): 82-116, 
130-184, 201-258, 240-279. The pointing and focus were always checked on 
the target source, since CRL618 exhibits a continuum of about 0.3 K in 
the T$_{A}^{*}$ scale of the IRAM-30m telescope, across the surveyed 
frequency range (see figure \ref{Fig1}). 
The pointing was kept within 2'' accuracy. We carried out the observations 
using the wobbler-switching mode with offsets of 60'' and 
frequencies of 1 Hz, in order to obtain very flat baselines. 
This is very important because some spectra are so crowded with 
lines that their reduction could be difficult otherwise (not enough baseline). 
The backends used were two 512 MHz filter banks connected to the receivers 
operating below 200 GHz, and a 512 MHz autocorrelator with channel width of 
1.25 MHz connected to the others. System 
temperatures were typically 100-400 K at 3 mm, 200-600 K at 2 mm 
and 300-800 K at 1.3 mm.

The observations performed in several observing runs in 2000, 2001 and 2002 
with the 10.4 meter dish of the Caltech Submillimeter Observatory (CSO) 
at the summit of Mauna Kea (Hawaii) made use of a helium-cooled SIS 
receiver operating in double-sideband mode in the frequency range 
280-360 GHz. The continuum emission from CRL618 at the CSO has a 
level of 0.05-0.06 K in the T$_{A}^{*}$ scale, too low to allow 
pointing on the source. The 
pointing was therefore checked on the available planets and was kept 
within 3-4'' accuracy (the HPBW ranged from 20'' to 25'').  The 
system temperatures varied from 500 K to 1200 K. In order to obtain 
flat baselines we used the chopping of the secondary mirror technique at a 
frequency of about 1 Hz and offset of 90''. In order to discriminate 
whether the lines are from the signal or from the image sideband, 
the observations were made twice with a frequency shift of a few 
tens of MHz (usually 50).

\section{HC$_{3}$N spectroscopy}
The rotational spectrum of cyanoacetylene, HC$_{3}$N, has been investigated 
in vibrational states up to about 1750 cm$^{-1}$ by Mbosei et al. 
(2000). The resulting data base for this molecule has been provided to 
us by A. Fayt. This work has been recently published (Fayt et al. 2004).   
the frequencies have been checked against our millimeter wave line survey 
with a remarkable agreement in the detected vibrational states.

There are three doubly degenerate bending modes of HC$_{3}$N, 
called $\nu_5$, $\nu_6$, and $\nu_7$ in order of decreasing energy 
(223, 499 and 663 cm$^{-1}$).  When considering 
the rotation, their degeneracy is broken due to the direction of the rotation axis. 
This is known as $\ell$-doubling. $\nu_4$ is the lowest energy stretching mode 
(880.6 cm$^{-1}$). The other 3 vibrational (stretching) modes have 
energies above 2000 cm$^{-1}$ and are not seen in CRL618. In this work, 
we have been able to detect HC$_{3}$N pure rotational lines in the 
ground and 14 different vibrationally excited states. 
Detections have been also obtained in a number of isotopomers, but they 
will be discussed in a separate work. 

The lines are labeled by the rotational quantum numbers (J$_{up}$--J$_{low}$), 
the vibrational quantum numbers ($v_4$,$v_5$,$v_6$,$v_7$) and 
the $\ell$-doubling parameters: [ $\ell_{5}$ $\ell_{6}$ $\ell_{7}$ ]. 
Table \ref{T1} provides the observed line parameters (velocity, width and integrated 
area) of a pair of rotational transitions (J=12--11 and J=26--25) for all detected 
vibrationally excited states. The J=12--11 lines always display an absorption feature 
centered at $\sim$ -27.5 kms$^{-1}$ accompanied by 
weaker emission (undetected in some cases) centered around the 
LSR velocity of the source. In the J=26--25 lines the absorption is barely 
detected and the emission is always much stronger. The selected rotational 
numbers are representative of the two different types of line profiles observed  
in HC$_{3}$N and in other molecules toward CRL618.

\section{Results}
\label{sct:results}
 
Of the complete frequency survey, as described in section \ref{sct:obs}, 
the data set used for the analysis presented in this paper consists of:

\begin{itemize}
\item The continuum measured in each setting across the frequency 
ranges covered: 80-116.5, 132-177 and 204-276 GHz.

\item Those spectra containing HC$_{3}$N rotational lines in a subset of 
vibrational states, selected to sample the range of the vibrational energies 
among those detected. The selected vibrational states are: (0000), (0001), 
(0002), (0010), (0100), (0003), (0011), (1000), (0012), and (0102) which 
vibrational energies are: 0, 223, 446, 499, 663, 669, 722, 881, 945, and 
1109 cm$^{-1}$ respectively.
\end{itemize}

\subsection{Continuum emission}
\label{sectcont}

The central star has an effective temperature of 30000 K and it is sourrounded 
by an HII region and its PDR. Continuum emission from the HII region and the dust 
affect the population of the molecular levels and has to be taken into account in 
modelling the source. The properties of the measured continuum will be discussed 
in this section.

\begin{figure}[t]
   \centering
 \includegraphics[angle=-90,width=\textwidth]{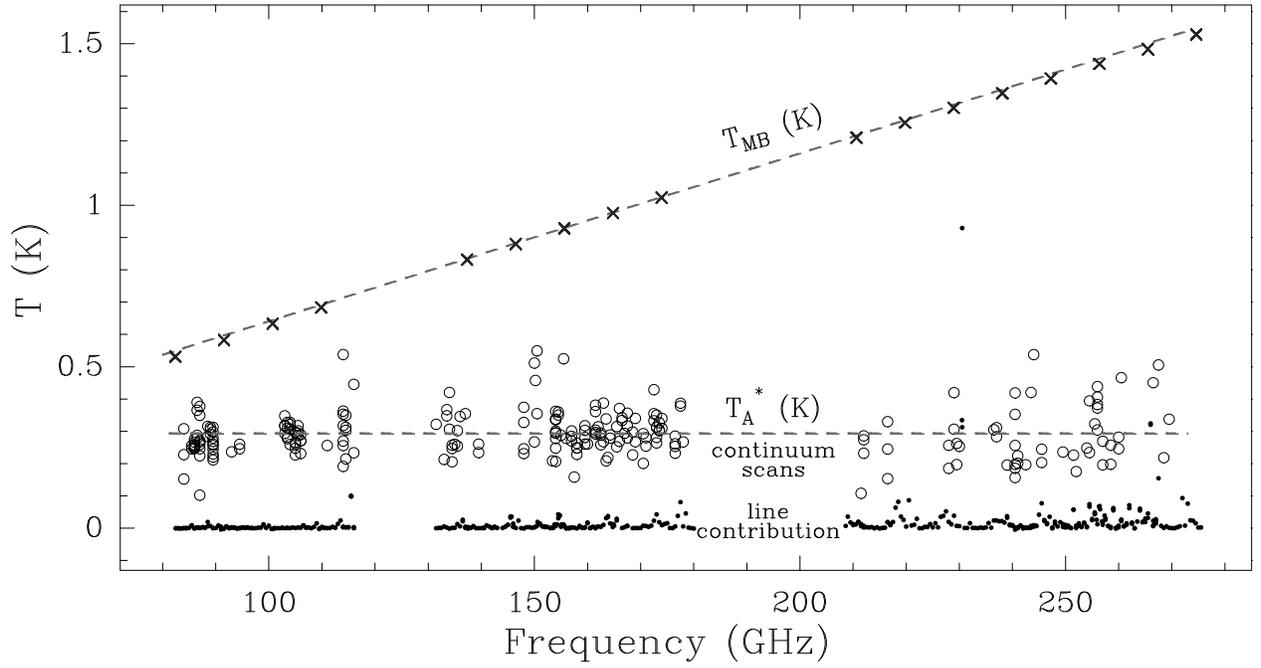}
   \caption{Open circles: Millimeter-wave continuum of CRL618 observed by the 
IRAM 30m telescope (T$_{A}^{*}$ scale, from pointing
 observations on CRL618 in continuum mode). Horizontal dashed line: Fit of these 
data to a horizontal line. Inclined dashed line: Fit translated into T$_{MB}$. 
Crosses: Simulated continuum T$_{MB}$ by our model (see sections \ref{sc:model} and 
\ref{sc:simulations}).}
              \label{Fig1}%
    \end{figure}

The level of continuum emission is important to explain the features 
seen in the line profiles of many  molecules in CRL618. Although 
the wobbler-switching method used at the 30m telescope for this work
should provide in principle a precise continuum level for each spectrum 
directly, the continuum levels of different spectra never match perfectly. 
Moreover, earlier work has shown evidence of variability over several years    
(see below). Therefore, 
we have decided to use all the pointing data to retrieve the continuum by fitting 
gaussians to the pointing scans and storing the peak of the fitted gaussian 
(in  T$_{A}^{*}$ scale) and the central frequency. The obvious bad scans and 
those where the pointing errors were larger than 2'' have been ignored. The 
results are shown in figure \ref{Fig1}(open circles). T$_{MB}$ and fluxes 
have been derived from a linear fit to those data (actually a horizontal 
line), scaled by the corresponding beam and forward efficiencies of the 30m 
telescope. In principle, this analysis could be affected by the presence of lines. 
However, the total flux from the lines represents 
less than 3-5 \% of to the continuum in most frequency 
settings. We have just ignored the few frequency settings 
with very strong lines such as the lowest rotational transitions of CO and HCN. In 
total flux our measurements give 1.75 Jy at 90 GHz, 2.4 Jy at 150 GHz and 3.4 Jy 
at 240 GHz. 


Several continuum measurements of CRL618 at different frequencies exist. In an 
early interferometric mapping carried out at cm wavelengths with the VLA it was 
shown that the compact radio source had a size of 0.4''$\times$0.1'' 
(Kwok \& Bignell 1984). Mart\'{\i}n-Pintado et al. (1993, 1995) 
using the VLA at 23 GHz found an elliptical source with size 0.40''$\times$0.12''  
and integrated flux density of 189$\pm$20 mJy (peak continuum flux:  
16.7 mJy~beam$^{-1}$) in 1993 and 250$\pm$20 mJy in 1995. Thorwirth et al. (2003) 
found 26.3$\pm$3 mJy at 4488.48 MHZ and 750$\pm$75 mJy at 40 GHz with a size of 
0.34''$\times$0.16'' (peak continuum flux: 140 mJy~beam$^{-1}$). At 1.3 mm, Walmsley 
et al. (1991) found fluxes varying from 1.0 to 3.3 Jy between 1987 and 1990. 
If this variability is real, it could explain part of the scattering in the data 
points presented in figure \ref{Fig1} (obtained over several winters). 
Earlier work (Kwok \& Feldman 1981, Spergel et al. 1983) also present the 
source as variable at 
centimeter wavelengths. Other measured fluxes found in the literature include Shibata 
et al. (1993), 1.48 Jy at 115 GHz, and Mart\'{\i}n-Pintado et al. (1998) who measured the 
following at 22.3, 87.6, 147.0 and 231.9 GHz: 0.24$\pm$0.02, 1.1$\pm$0.1, 1.8$\pm$0.2, 
2.0$\pm$0.3 Jy, values about 25-32 \% smaller than our own results 
(see figure \ref{Fig1}) that represent an average over at least 4 years.


From the line survey currently under way in the 280-360 GHz range, 
the measured continuum emission averages 0.06-0.07 K in the T$_{A}^{*}$ 
scale ($\sim$ 0.1 K in T$_{MB}$). This result agrees with the trend 
observed at the higher end of the IRAM 30m survey (275 GHz) given the 
smaller size of the CSO antenna (10.4 m).

\subsection{HC$_{3}$N ground vibrational state}

\begin{figure}[t] 
   \centering
\includegraphics[width=7.0cm]{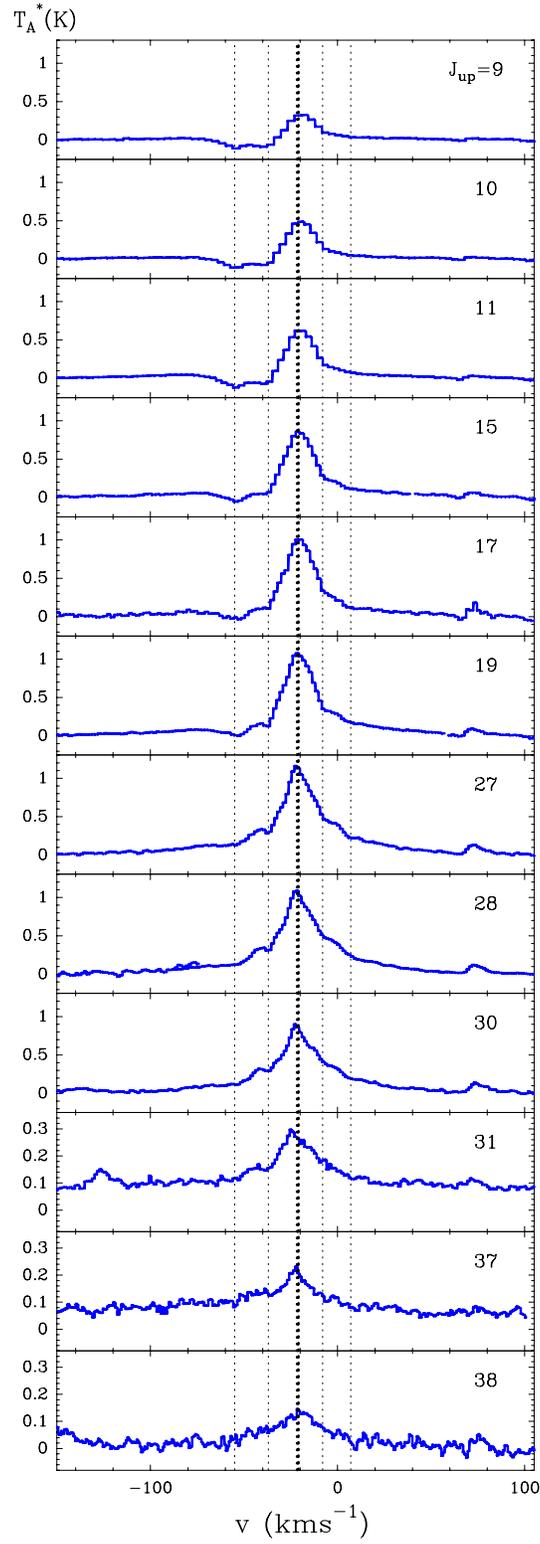}
\caption{HC$_{3}$N pure rotational lines toward the proto-planetary 
nebula CRL618 in the ground vibrational state. The thick dotted line marks the 
LSR velocity of the source. The thin dotted lines mark interesting features in 
the line profile discussed in the text.}
\label{Fig2}%
\end{figure}

\begin{figure}[t]
   \centering
\includegraphics[width=7.5cm]{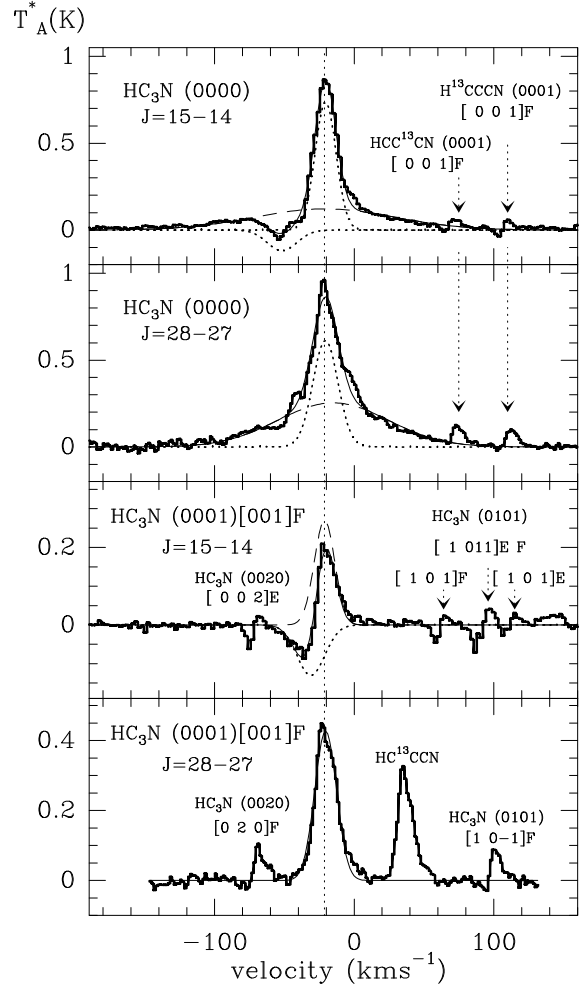}
\caption{Velocity components of HC$_{3}$N lines toward CRL618. In the 
ground vibrational state (top two spectra) we see broad wings associated 
to the HVMW discovered by Cernicharo et al. 1989 and a narrower emission  
due to the slowly expanding gas envelope around the HII region. A blending with 
HC$_{3}$N (0100) causes the absorption feature at $\sim$-37 kms$^{-1}$. The 
absorption centered at -55 kms$^{-1}$ is also seen in other abundant molecules 
such as HCN and CO. In vibrationally excited states (bottom two spectra) the 
HVMW component is not visible. At low-J the absorption of the slowly expanding 
gas in front of the continuum source is seen centered at $\sim$-27 kms$^{-1}$. However, 
the absorption dissapears at J $>$ 20 for the particular case of the (0001) state (the 
J number where this happens slightly depends on the vibrational state). Still, rotational 
lines within the (0001) state trace larger velocities than those in higher vibrational 
states, although it is not the HVMW.} 
\label{Fig3}%
\end{figure}

The profiles of HC$_{3}$N rotational lines within the ground 
vibrational state (see figure \ref{Fig2}) show evidence of the high 
velocity molecular wind (HVMW), also seen in lines of CO and other molecules. 
In HC$_{3}$N, the HVMW component 
of the line profile extends some 200 kms$^{-1}$ in some cases (see figure 
\ref{Fig3}). In addition to this one, there is another component associated 
to a slowly expanding molecular envelope around the ultracompact HII 
region. This component has a maximum half power width of about 18 
kms$^{-1}$ and is approximately centered at the LSR 
velocity of CRL618 (-21.3 kms$^{-1}$). Two different absorption 
dips are present (see figure \ref{Fig2}). The first one appears at 
-37 kms$^{-1}$ in all the lines at least up to 
J=37--36. It turns out that this feature is related to the [ 1 0 0]E 
component of the rotational transitions of the (0100) state. 
The other absorption feature at -55 kms$^{-1}$ is due to expanding gas in the 
line of sight of the continuum source and is also found in 
low-J HCN (Neri et al. 1992) and CO (Cernicharo et al. 1989, 
Herpin et al. 2002). On the red side of 
the line, between approximately -8 and 7 kms$^{-1}$, there is evidence of emission that 
Neri et al. 1992 attributed to an unresolved structure located to the North-East of the HII region.

\subsection{Vibrationally excited HC$_{3}$N}

\subsubsection{The (0001) vibrational state}
The HC$_{3}$N rotational lines in the (0001) vibrational level, at 
least below J$_{up}$ $\sim$ 20, still trace gas at higher velocities than 
that seen in rotational lines corresponding to higher energy vibrational 
levels (see below). For example, the J=15-14   
transition in (0001) shows an absorption dip that clearly starts below 
-50 kms$^{-1}$, more than 30 kms$^{-1}$ apart from the LSR velocity of the 
source (see figure \ref{Fig3}). However, for J$_{up}$ above 20 this colder and 
faster outflow is not significant in the level population anymore. 

\subsubsection{Other vibrational states} The HC$_{3}$N lines in the rest of 
the detected vibrationally excited states display patterns with many common 
characteristics. The 
lines appear dominantly in emission at 1.3 mm (at the IRAM-30m  radiotelescope) and 
at shorter wavelengths (CSO observations at 0.850 mm) with a half power width ranging from 5 to 
16 kms$^{-1}$, but mostly around 8-10 kms$^{-1}$. However, the majority of lines 
in the 2 and 3 mm windows show P-Cygni profiles, with the absorption part being 
systematically deeper at lower frequencies. This absorption is centered in the range 
-26 to -31 kms$^{-1}$, as expected for an expanding envelope with velocities around 
5-10 kms$^{-1}$ surrounding the central continuum source at the LSR velocity of 
CRL618. The absorption is less 
deep and shifts to more negative velocities as the emission part gets stronger with 
increasing J. At frequencies well below 80 GHz, almost all lines, from other 
species also, appear strictly in absorption.
The velocity of the absorption feature in HC$_{3}$N 
rotational lines, with J$_{up}$ $<$ 7, in different vibrationally 
excited states, observed with the Effelsberg 100m radiotelescope 
(Wyrowski et al. 2003) is around -26 or -27 kms$^{-1}$. 
Thorwirth et al. 2003 also report HCN $\Delta$J=0, $\ell$-doubling            
transitions purely in absorption at the same velocity, as do Mart\'{\i}n-Pintado 
et al. (1995) for NH$_{3}$ (3,3). The lowest 
transitions we have observed (J=9-8, J=10-9) in some highly excited 
vibrational states such as (0100) and (1000) do not show emission 
and the absorption feature is also centered around -27.5 kms$^{-1}$. 




\section{Discussion}

In this work we assume that the bulk of vibrationally excited HC$_{3}$N 
emission comes from a compact envelope expanding at about 10 kms$^{-1}$ around 
the central continuum source. The availability of rotational ladders in so many 
different vibrationally excited HC$_{3}$N states 
should provide a quite precise picture of the size ratio 
of the slowly expanding gas region with respect to the continuum source, the shape 
of the envelope, the temperature profile, velocity field and 
HC$_{3}$N density distribution. All this is explored in 
sections \ref{sc:model} and \ref{sc:simulations}.

\subsection{Model}
\label{sc:model}
In order to reproduce the observed rotational HC$_{3}$N ladders we have built 
a model, for LTE conditions, of an expanding envelope   
around a continuum source. Because of blending with the 
selected HC$_{3}$N lines for the analysis, we have also introduced HC$_{5}$N with an 
abundance ratio of 1/3 respect to HC$_{3}$N (Cernicharo et al. 1987; Gu\'elin and 
Cernicharo 1991), and isotopic species of 
HC$_{3}$N using a $^{12}$C/$^{13}$C ratio of 15. 

The lines are formed as follows: First, the central continuum source is responsible 
for the observed continuum level. The gas situated between the continuum 
source and the Earth absorbs that continuum and also emits. 
The gas in regions outside the line of sight to the continuum source contributes by 
emission only. The continuum source is considered opaque to the emission of the 
gas behind it. The 
spectrum of the continuum source and its size ratio with respect to the gas region, 
turn out to be key parameters to fit the simulations to the observed HC$_{3}$N rotational 
lines. The absorption takes place primarily at the terminal velocity of the 
expanding envelope in front of the continuum source. The emission can help 
to ``fill'' the absorption dip depending on the characteristics of the velocity 
field and the turbulence velocity.

\begin{figure}[t]
   \centering
\includegraphics[width=9.5cm]{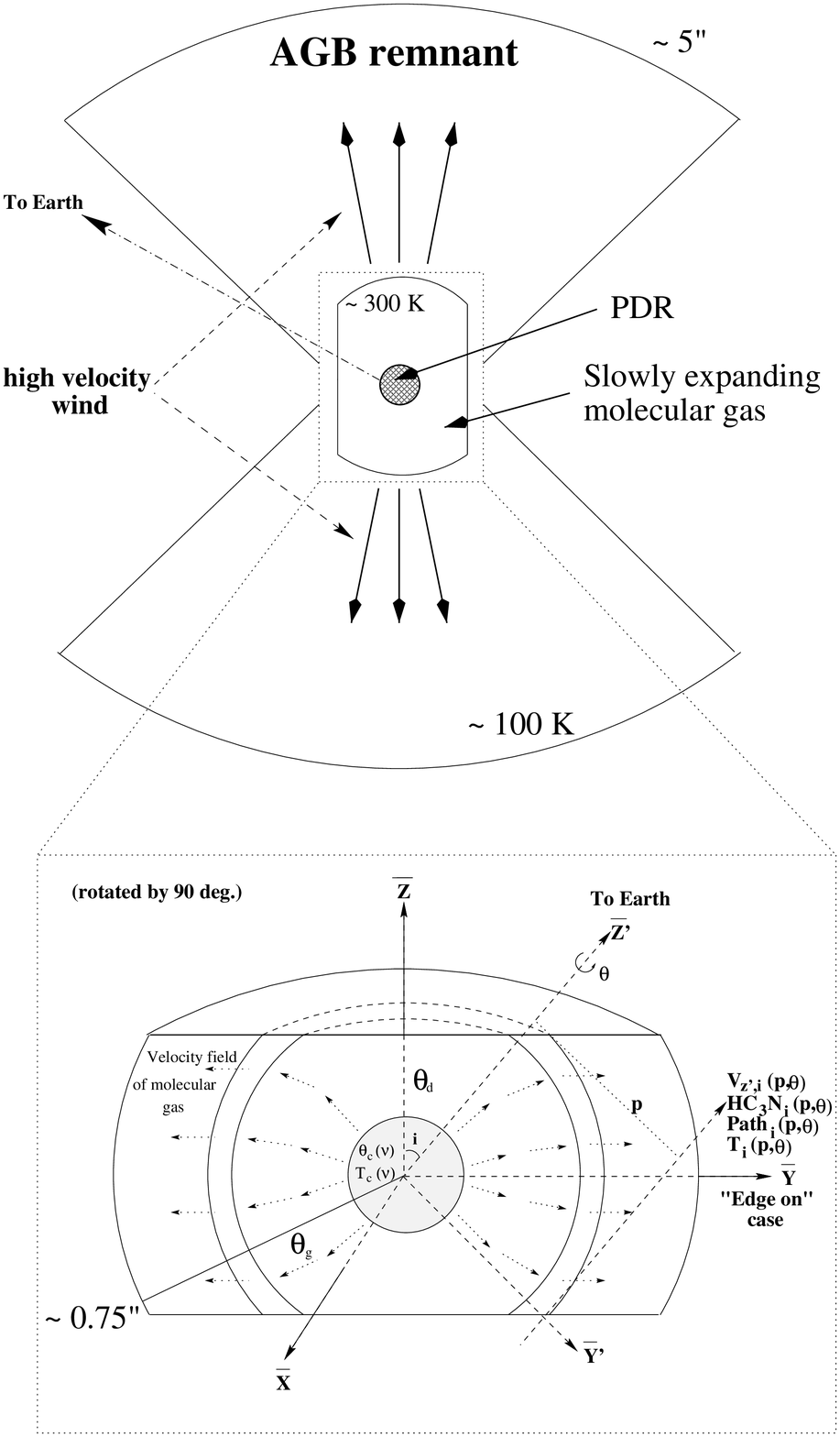} 
   \caption{Model of CRL618 including the description of the expanding elongated structure 
considered to account for the observed HC$_{3}$N line profiles. See text for details about 
the different parameters of the model.} 
              \label{Fig4}%
    \end{figure}

The simplest model considers a spherical expanding envelope of gas  
surrounding a central continuum source (also spherical). The envelope is 
divided into different shells where the total HC$_{3}$N number density 
and the temperature vary according to a power law, and radial expansion and turbulent 
velocities are  linearly interpolated between input values at the boundaries. 
Such a simple model can only approximately reproduce the main observational 
characteristics of the  HC$_{3}$N rotational ladders in CRL618. A more complex 
model has been developed to account for an elongated envelope with non-radial 
velocity components. We consider an external angular size of the (spherical) 
slowly expanding envelope $\theta_{g}$ and 
then we truncate it so that, in the truncation direction, the size is: 
$\theta_{d}$=$\theta_{g}\times r_{d}$, $r_{d}$ being the truncation parameter  
(see figure \ref{Fig4}). Other geometric details of this model are the following: 
A first coordinate system (XYZ) is defined with the Z axis running in the 
direction where the sphere is truncated, and X, Y are orthogonal to each other and 
to Z. The velocity field will be a function of 
[(x$^2$+y$^2$)$^{\frac{1}{2}}$,z] (azimuthal symmetry in the XY plane), and 
the temperature and gas densities will only be a function 
of r=(x$^2$+y$^2$+z$^2$)$^{\frac{1}{2}}$ (radial symmetry). The second 
coordinate system has the axis X' in common with X just described, 
and the remaining two (Y',Z') are rotated respect to (Y,Z) by an angle 
$i$ so that Z' points to the Earth (line of sight). The ``edge-on'' case 
therefore corresponds to i=90$^{o}$. The continuum 
source is spherical with an angular radius $\theta_{c}$. 
The value of $\theta_{c}$ has to be in the range 0.15''-0.4'' in order to be 
compatible with data presented in section \ref{sectcont}. We use as input 
parameter its ratio $r_{c}$ respect to the external angular size of the 
slowly expanding envelope, $r_{c}$=$\theta_{c}$/$\theta_{g}$. The 
effective temperature of the continuum source would depend on that size 
in order to match the observed continuum. In general, the 
equivalent size $\theta_{c}$ would be a function of frequency. However, 
a constant size and a frequency dependent $T_{c}$, characterized by a  
spectral index $s$ (see following equation), fit the data very nicely 
(see figure \ref{Fig1} and table \ref{tablecont}). 

\begin{equation}
T_{cont}(\nu)=T_{cont}({\rm 200\;\;\;GHz})[\nu/200]^{s}{\rm ,} 
\label{tcont}
\end{equation}

\begin{figure}[t]
   \centering
   \includegraphics[width=10cm,angle=270]{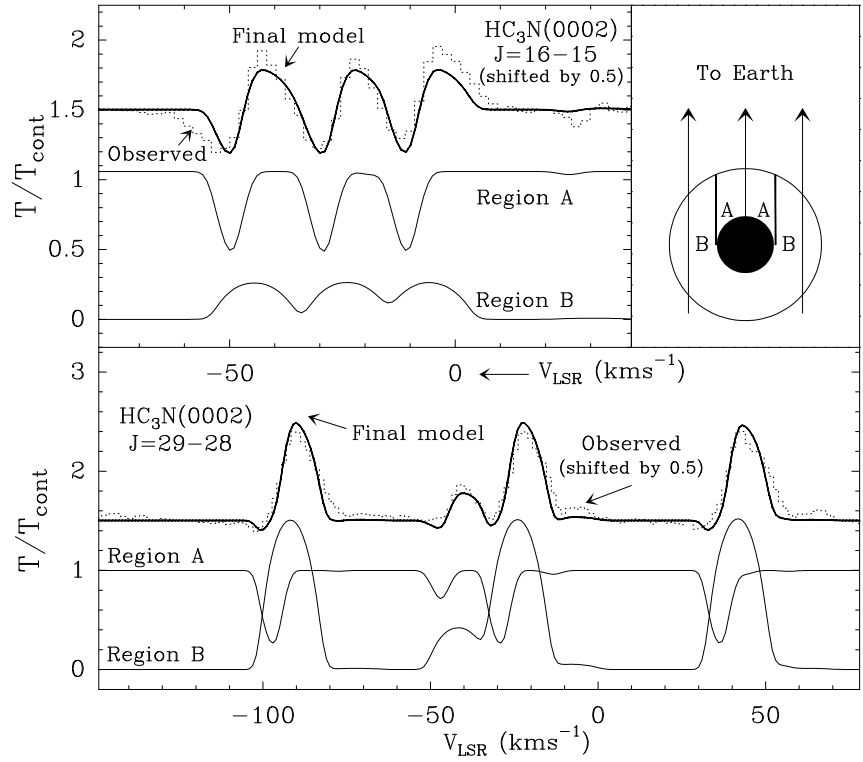}
\caption{Fit of observed HC$_{3}$N rotational line toward CRL618 and 
separate contribution from gas in and out of the line of sight to the 
continuum source. The physical parameters correspond to those listed in 
table \ref{table2}.} 
\label{Fig5}%
\end{figure}

In the elongated envelope case, the velocity field cannot be exactly radial: 
$\vec{v}=v_{r}\vec{r}+v_{xy}\vec{r_{xy}}$ ($v_{xy}\ne 0$ if r$_{d}<$1). The extra 
velocity component in the XY 
plane makes the 
velocity pattern look as in figure \ref{Fig4}. This kind of velocity field has 
the advantage of having gas with the same velocity projection in regions in and out 
the column between the continuum source and the Earth. Some extra gas is thus 
allowed to contribute filling the absorption dip (not only the turbulence velocity 
helps to do this, see figure \ref{Fig5}). The velocity field remains 
unchanged by rotation respect to the Y or Z axis. Therefore, in the $i$=0$^{o}$,90$^{o}$ 
cases (``edge-on'' or ``orthogonal'' envelope cases), only the radiative transfer results 
as a function of the impact parameter $p$ (defined as an angle) have to be used in the 
numerical convolution with the antenna beam pattern. If $i$ takes any intermediate 
value, the numerical integration becomes dramatically more time consuming (10-100 times) 
because the azimuthal symmetry respect to the XY' plane is broken and 
the convolution has to be 
performed in both $p$ and $\theta_{XY'}$ (azimuthal angle around 
the Z' or line-of-sight axis, see figure \ref{Fig4}).  

As we have seen, the model has an important number of parameters. Some of them can be 
fixed with some confidence, based on data such as the continuum flux (see figure \ref{Fig1}) 
and the estimated distance to the source. The other parameters, velocity 
field, kinetic temperature, shape of the slowly expanding envelope, ratio 
between its size and that of the continuum source, and HC$_{3}$N column 
density, can be constrained by running a number of models. It is quite  
straightforward to find the best values for all parameters, with the exception 
of the kinetic temperature and HC$_{3}$N column density. This is discussed below.





\subsection{Constraining the physical parameters}
\label{sc:simulations}
The first step in the simulation process has concentrated on the continuum source. We have 
considered different sizes for it (constant with frequency) and we have calculated 
its effective temperature at a reference frequency (200 GHz) and the spectral index, $s$, to 
match the observed continuum (figure \ref{Fig1}). Based on the references given 
in section \ref{sectcont}, the range of sizes we have explored is 0.15'' to 0.40''. The 
results are given in table \ref{tablecont}. For each set of values, the 
fit to our continuum data is very good, typically as shown on figure \ref{Fig1}. As the 
external size of the gas region we are modeling is around 1.5'' and the ratio of this size to 
the continuum source size is a key parameter for reproducing the observed line profiles, the 
preferred values among the five given alternatives are highlighted in bold face.

 As the number of physical parameters 
and lines observed are both large, a general numerical fit to constrain all physical 
parameters simultaneously is not possible. Rather, it is necessary to run a large 
number of models in several steps in order to find a reasonable match to all the 
observed HC$_{3}$N rotational ladders. As this is very time consuming, the 
simulations have been restricted to the (0002), (0010), (0100), (0003), (0011), (1000), 
(0012), and (0102) vibrational states, as explained in section \ref{sct:results}.

\begin{figure*}[t]
   \centering
   \includegraphics[width=15cm]{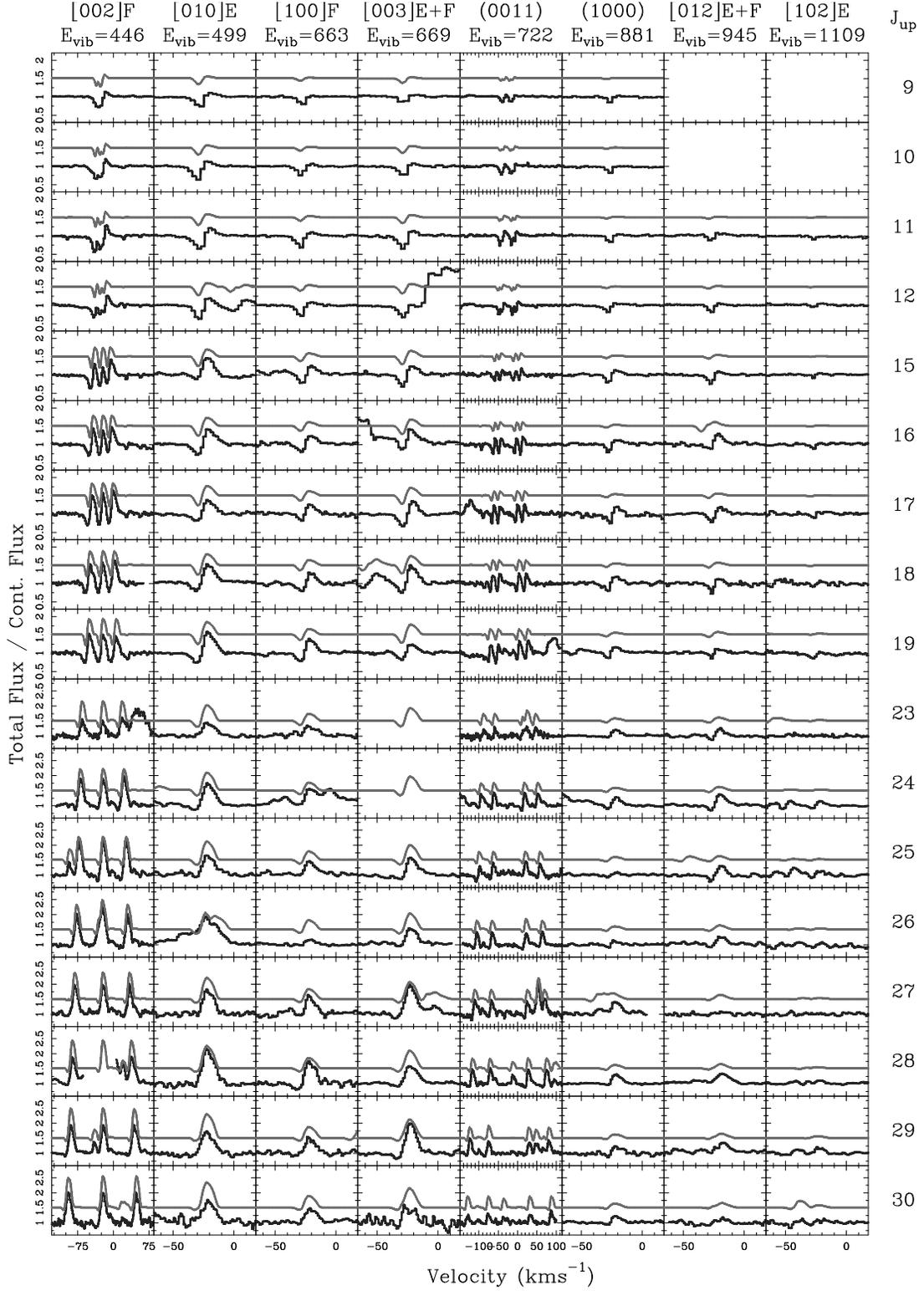}
   \caption{Observed HC$_{3}$N pure rotational lines toward the proto-planetary 
nebula CRL618 in several vibrationally excited states from 446 to 1109 cm$^{-1}$ compared 
to the results of our model. In a few cases the observation is not shown because the corresponding 
HC$_{3}$N line is totally masked by a stronger line from another molecule. }
              \label{Fig6}%
    \end{figure*}

In the search for the best solution we also need to define a parameter to evaluate 
the quality of the fit. Ideally, this parameter has to give equal weight to the different 
lines, independently of the frequency resolution. We thus defined an overall $\chi$ 
as follows:

\begin{equation}
\chi=\frac{1}{N_{l}} \sum_{i} \sqrt{\frac{\sum_{j}|Yd-Ym|^{2}}{N_{ch}}} 
\label{chi}
\end{equation} 
where $N_{l}$ is the number of suitable lines in figure \ref{Fig6}, $Yd(j)$ are the 
data (total flux divided by the contiuum flux), $Ym(j)$ are the model results, 
$N_{ch}$ is the number of channels where $Ym(j)\ne 1.0$, 
i.e. we do not consider channels where the signal is just the continuum level, or lines 
from species other that HC$_3$N or HC$_5$N. In addition, if a line from other chemical 
species overlaps with those from HC$_3$N or HC$_5$N, the corresponding spectrum is 
discarded. Note also that in figure \ref{Fig6} not all the $\ell$-type lines 
are shown in each case due to the sometimes large velocity separation. Only the data 
shown in the figure are used to calculate $\chi$. Blending between different 
vibrational states or within the same vibrational state has been treated in 
our simulations. 

\begin{figure}[t]
   \centering
   \includegraphics[width=10cm]{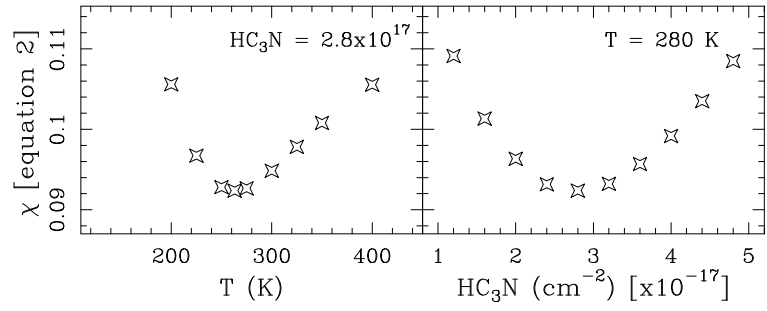}
   \caption{Value of $\chi$ (defined in equation \ref{chi}) for the grid of models 
run to find the temperature of the CRL618 slowly expanding envelope and the HC$_3$N 
column density that best fits the data shown on figure \ref{Fig6}.}
              \label{Fig7}%
    \end{figure}

The two parameters that we have tried to determine by looking at $\chi$ in a grid of 
models are the temperature of the gas and the HC$_3$N column density. In 
order to illustrate how the best solution (\ref{Fig6}) is found we show in figure \ref{Fig7} 
the values of parameter $\chi$ as a function of these two 
parameters. The overall set of parameters characterizing the slowly expanding 
molecular envelope is then shown on  table \ref{table2}.



The kinetic temperature found here (250-275 K) is in very good agreement with the 
value derived from ISO IR data by Cernicharo et al. 2001a,b. 
The average column density of absorbing HC$_{3}$N in front of 
the continuum source is found to be in the range 2.0-3.5$\cdot$10$^{17}$ cm$^{-2}$. 
HC$_{3}$N column densities 
in front of the continuum source where determined to be 5$\cdot$10$^{16}$ 
cm$^{-2}$ by Cernicharo et al. (2001a) from ISO observations of HC$_{3}$N 
(0100) and (0001) bending modes in absorption around 14 $\mu$m. 
The difference between both estimates could be related to two facts: First, 
the ISO data lacks the spectral resolution needed to fully resolve the individual 
lines used for the estimate. Second, Cernicharo et al. 2001a assumed the same 
filling factor for the continuum source and the absorbing gas. However, dust emission 
at the wavelength of the (0100) mode of HC$_3$N, and in particular at that 
of the (0001) mode could contribute from a region larger than the 
zone where HC$_3$N is produced. 
 The IR ISO data could give the same HC$_{3}$N column density as that 
derived here from the millimeter lines, if the 
continuum level to be considered for the former calculation (originated from 
behind the gas) is only about 30-50 \% of the total observed continuum 
flux at 14 $\mu$m.

\section{Conclusions}
In this paper we have shown that toward the protoplanetary 
nebula CRL 618:

\begin{itemize}
\item HC$_{3}$N is detected in its ground and other vibrational states 
 with energies up to 1100 cm$^{-1}$. The J$_{up}$ range surveyed is from 9 to 30 
(IRAM 30m telescope, 15 vibrational states detected) and 31 to 
39 (CSO telescope, detections in 5, possibly 6 vibrational states). The observed 
frequencies match very well those provided by the model of Fayt et al. (2004).

\item The line profiles of the HC$_{3}$N ground vibrational state show evidence 
of the HVMW already found in the strong lines of CO, HCN and HCO$^{+}$ with the 
same velocity span ($\sim$200 kms$^{-1}$). Other vibrational 
states do not show this component in their rotational lines, except for the 
(0001) state that shows it marginally in the lowest J lines. 
Absorption features 
in the rotational lines of the ground vibrational state appear at similar velocities 
to those of features already seen in CO and HCN. The only exception is the one 
centered at $\sim$-37 kms$^{-1}$ that is due to blending with (0100)[1 0 0]E lines.

\item The pure rotational transitions in vibrationally excited states of HC$_{3}$N 
show an evolution in their line profiles with increasing J going from almost pure 
absorption centered at around -27 or -28 kms$^{-1}$, to P-Cygni profiles and, 
finally pure emission centered quite close to the LSR velocity of 
the source. The intermediate-J P-Cygni profiles have the emission part also centered 
quite close to the LSR velocity, while the minimum of the absorption part shifts to 
more negative velocities and becomes less deep as J increases. The exact value of 
J$_{up}$ at which the absorption disappears changes only slightly with the vibrational 
state: It happens at slightly higher values as the energy of the vibrational 
state increases. For example, no absorption is observed in the J=24-23 transition of the 
(0001) vibrational state, whereas some absorption is still 
seen in the J=26-25 line of the (1000) one. Some rotational lines within the 
(0001) state still trace larger velocities than those in the other vibrationally 
excited states.

\item An model for the slowly expanding envelope, under LTE, built 
to explore the physical parameters that explain the extensive set 
of HC$_{3}$N rotational lines detected ($\sim$300, although the fit has 
been done to a subset of 116) has shown that:  
\begin{enumerate}
\item The size of the inner continuum source is 
$\sim$0.27'' with an effective temperature of $\sim$3900 K at 200 GHz and a spectral index of -1.12. 

\item The external size of this slowly expanding envelope is $\sim$ $\theta$=1.5''.

\item The expansion velocity field has a radial component ranging from 5 to 12 kms$^{-1}$ with 
a possible extra azimuthal component reaching 6 kms$^{-1}$ at $\theta$=1.5''.

\item The turbulence velocity is $\sim$3.5 kms$^{-1}$. 

\item The temperature of the envelope is in the range 250-275 K.  

\item The HC$_{3}$N column density in front of the continuum source is in the range 
2.0-3.5$\cdot$10$^{17}$ cm$^{-2}$ (the best fit is obtained for 2.8$\cdot$10$^{17}$ cm$^{-2}$). 
\end{enumerate}

 \end{itemize}

\acknowledgments
The authors are grateful to the IRAM-30m staff for providing assistance 
during the observations. This work has 
been supported by NSF grant \# ATM-9616766, and by 
Spanish DGES and PNIE grants ESP2002-01627, AYA2002-10113-E and 
AYA2003-02785-E. CSO operations were supported by NSF grant AST-9980846.

\begin{table*}
\caption[]{Observed line parameters of the J=12-11 and J=26-25 transitions in all vibrationally excited 
states of HC$_{3}$N (see text for notation) detected toward CRL618 (v$_{1,2}$: position in 
kms$^{-1}$, $\Delta$v$_{1,2}$ 
width in kms$^{-1}$, A$_{1,2}$ area in K$\cdot$km$^{-1}$; 1 refers to the emission part and 2 to 
the absorption part of the line profile). In addition to the vibrational states show here, the (0111) 
at more than 1200 cm$^{-1}$ is tentatively detected.}
\label{T1}
\begin{center}
\begin{tabular}{|lr|ccccccc|ccccccc|}
\hline
{\footnotesize V. state} & {\footnotesize type of }& \multicolumn{7}{c|}{{\footnotesize J=12-11} }& \multicolumn{7}{c|}{{\footnotesize J=26-25}} \\
{\footnotesize {\bf  [E cm$^{-1}$]} }& {\footnotesize  $\ell$ transition }& {\footnotesize  $\nu$ (GHz) }& {\footnotesize  v$_{1}$ }& {\footnotesize $\Delta$v$_{1}$ }& {\footnotesize A$_{1}$ }& {\footnotesize   v$_{2}$ }& {\footnotesize $\Delta$v$_{2}$ }& {\footnotesize A$_{2}$ }& {\footnotesize }  
{\footnotesize v$_{1}$ }& {\footnotesize  $\nu$ (GHz) }&{\footnotesize $\Delta$v$_{1}$ }& {\footnotesize A$_{1}$ }& {\footnotesize   v$_{2}$ }& {\footnotesize $\Delta$v$_{2}$ }& {\footnotesize A$_{2}$ }\\
\hline
{\footnotesize (0001) }& {\footnotesize {[ 0 0 1]F} }&109.598825 &{\footnotesize -20.9 }& {\footnotesize 14.9 }& {\footnotesize 3.6 }& {\footnotesize  -31.6 }& {\footnotesize 22.3 }& {\footnotesize -3.3 }& 237.432241 &{\footnotesize -20.8 }& {\footnotesize 18.7 }& {\footnotesize 7.2 }& {\footnotesize }& {\footnotesize }& {\footnotesize  }\\
{\footnotesize   [223] }& {\footnotesize {[ 0 0 1]E} }&109.442011 & \multicolumn{6}{c|}{{\footnotesize blended with (0010)[ 0 1 0]F} }& 237.093375 & {\footnotesize }& {\footnotesize }& {\footnotesize }& {\footnotesize }& {\footnotesize }& {\footnotesize }\\
\hline
{\footnotesize  (0002) }&{{\footnotesize [ 0 0 0]E} }&109.865960 & \multicolumn{6}{c|}{{\footnotesize blended with [ 0 0 2]E,F} }& 237.968847 & {\footnotesize  -23.7 }& {\footnotesize 16.0 }& {\footnotesize 5.9 }& {\footnotesize }& {\footnotesize }& {\footnotesize  }\\
{\footnotesize   [426] }&{{\footnotesize [ 0 0 2]E} }&109.870288 & \multicolumn{6}{c|}{{\footnotesize blended with [002]F,[000]E} }& 238.053887  & {\footnotesize  -21.6 }& {\footnotesize 11.3 }& {\footnotesize 3.5 }& {\footnotesize }& {\footnotesize }& {\footnotesize }\\
{\footnotesize }& {\footnotesize {[ 0 0 2]F} }&109.865960 & \multicolumn{6}{c|}{{\footnotesize blended with [002]E,[000]F} }& 238.010133 &{\footnotesize -21.5 }& {\footnotesize 10.6 }& {\footnotesize 3.2 }& {\footnotesize }& {\footnotesize }& {\footnotesize  }\\
\hline
{\footnotesize (0010) }& {\footnotesize {[ 0 1 0]E} }& 109.352748&{\footnotesize  -21.3 }& {\footnotesize 7.4 }& {\footnotesize 0.47 }& {\footnotesize  -29.8 }& {\footnotesize 7.3 }& {\footnotesize -0.90 }& 236.900342 & {\footnotesize }& {\footnotesize }& {\footnotesize }& {\footnotesize }& {\footnotesize }& {\footnotesize       }\\
{\footnotesize [499] }& {\footnotesize {[ 0 1 0]F} }& 109.438718 &\multicolumn{6}{c|}{{\footnotesize blended with (0010)[ 0 1 0]F}  }& 237.086494 & {\footnotesize }& {\footnotesize }& {\footnotesize }& {\footnotesize }& {\footnotesize }& {\footnotesize    }\\
\hline
{\footnotesize (0100) }& {\footnotesize {[ 1 0 0]E} } & 109.183008 &\multicolumn{6}{c|}{{\footnotesize blended with ground state of HC$_{3}$N } } & 236.529566 &  \multicolumn{6}{c|}{{\footnotesize blended with ground state of HC$_{3}$N } } \\
{\footnotesize [663] }& {\footnotesize {[ 1 0 0]F} }& 109.244206 &{\footnotesize  -21.3 }& {\footnotesize 11.0 }& {\footnotesize 0.35 }& {\footnotesize  -28.4 }& {\footnotesize 6.6 }& {\footnotesize -0.66     }& 236.661402 & {\footnotesize   -20.5 }& {\footnotesize 8.4 }& {\footnotesize 0.44 }& {\footnotesize }& {\footnotesize }& {\footnotesize  }\\
\hline
{\footnotesize (0003) }& {\footnotesize {[ 0 0 1]E} }& 110.050809 &{\footnotesize  -21.3 }& {\footnotesize 9.0 }& {\footnotesize 0.37 }& {\footnotesize   -28.7 }& {\footnotesize 6.9 }& {\footnotesize -0.59 }& 238.401147 & {\footnotesize  -21.2 }& {\footnotesize 7.47 }& {\footnotesize 1.0 }& {\footnotesize }& {\footnotesize }& {\footnotesize  }\\ 
{\footnotesize [669] }& {\footnotesize {[003]E+F}  }& 110.211793 &{\footnotesize  -21.3 }& {\footnotesize 4.1 }& {\footnotesize 0.02 }& {\footnotesize   -29.5 }& {\footnotesize 7.7 }& {\footnotesize -0.79 }& 238.771331 & {\footnotesize  -21.3 }& {\footnotesize 12.0 }& {\footnotesize 1.8 }& {\footnotesize -31.1 }& {\footnotesize 5.8 }& {\footnotesize -0.3 }\\
{\footnotesize }& {\footnotesize {[ 0 0 1]F} }& 110.366468 & {\footnotesize  -21.3 }& {\footnotesize 14.6 }& {\footnotesize 0.7 }& {\footnotesize   -28.3 }& {\footnotesize 6.2 }& {\footnotesize -0.62 }& 239.082313 & {\footnotesize  -20.9 }& {\footnotesize 11.5 }& {\footnotesize 1.4 }& {\footnotesize }& {\footnotesize }& {\footnotesize  }\\
\hline
{\footnotesize (0011) }& {\footnotesize {[ 0 1-1]E} }& 109.736687 &\multicolumn{6}{c|}{{\footnotesize  $\sim$3 MHz away from [ 0 1-1]F}  }& 237.681318 & {\footnotesize  -22.1 }& {\footnotesize 9.2 }& {\footnotesize 1.5 }& {\footnotesize -28.2 }& {\footnotesize 14.2 }& {\footnotesize -0.7  }\\
{\footnotesize  [721] }& {\footnotesize {[ 0 1-1]F} }& 109.740137 & \multicolumn{6}{c|}{{\footnotesize  $\sim$3 MHz away from [ 0 1-1]E}  }& 237.713612 & \multicolumn{6}{c|}{{\footnotesize blended with (0101)[ 1 0 1]F} }\\
{\footnotesize }& {\footnotesize {[ 0 1 1]F} }& 109.749784 &\multicolumn{6}{c|}{{\footnotesize  $\sim$3 MHz away from [ 0 1 1]E}  }& 237.783921 & {\footnotesize  -21.8 }& {\footnotesize 9.9 }& {\footnotesize 1.9 }& {\footnotesize -26.9 }&{\footnotesize 15.5 }& {\footnotesize -1.0 }\\
{\footnotesize }& {\footnotesize {[ 0 1 1]E} }& 109.752621 &\multicolumn{6}{c|}{{\footnotesize  $\sim$3 MHz away from [ 0 1 1]F}  }& 237.814870 & {\footnotesize  -22.9 }& {\footnotesize 8.6 }& {\footnotesize 1.4 }& {\footnotesize -28.3 }& {\footnotesize 11.0 }& {\footnotesize -0.7 }\\
\hline
{\footnotesize (1000) }& {\footnotesize {[ 0 0 0]E} }& 109.023292 &{\footnotesize  -21.3 }& {\footnotesize 12.7 }& {\footnotesize 0.22 }& {\footnotesize  -27.9 }& {\footnotesize 6.2 }& {\footnotesize -0.40  }& 236.184052 & {\footnotesize  -23.6 }& {\footnotesize 8.8 }& {\footnotesize 0.35 }& {\footnotesize  -27.8 }& {\footnotesize 4.1 }& {\footnotesize -0.22 }\\
{\footnotesize [880] }& {\footnotesize }& {\footnotesize }& & {\footnotesize }& {\footnotesize }& {\footnotesize }& {\footnotesize }& {\footnotesize }& {\footnotesize }& & {\footnotesize }& {\footnotesize }& {\footnotesize }& {\footnotesize            }& {\footnotesize     }\\
\hline
{\footnotesize (0101) }& {\footnotesize {[ 1 0-1]F} }& 109.558099 &{\footnotesize  -21.3 }& {\footnotesize 6.0 }& {\footnotesize 0.07 }& {\footnotesize  -28.0 }& {\footnotesize 4.5 }& {\footnotesize -0.25  }& 237.336037 & {\footnotesize   -21.3 }& {\footnotesize 12.0 }& {\footnotesize 0.73 }& {\footnotesize  -29.0 }& {\footnotesize 5.2  }& {\footnotesize -0.30   }\\
{\footnotesize [886] }& {\footnotesize {[ 1 0-1]E} }& 109.549318 &{\footnotesize  -21.3 }& {\footnotesize 7.5 }& {\footnotesize 0.16 }& {\footnotesize  -27.5 }& {\footnotesize 3.9 }& {\footnotesize -0.18  }& 237.409821 & \multicolumn{6}{c|}{{\footnotesize  blended with (0001)[ 0 0 1]F}   }\\
{\footnotesize }& {\footnotesize {[ 1 0 1]F} }& 109.564687 &{\footnotesize  -21.3 }& {\footnotesize 7.5 }& {\footnotesize 0.16 }& {\footnotesize  -27.5 }& {\footnotesize 5.5 }& {\footnotesize -0.29  }& 237.716883 & \multicolumn{6}{c|}{{\footnotesize  blended with (0011)[ 0 1-1]F}     }\\
{\footnotesize }& {\footnotesize {[ 1 0 1]E} }& 109.553091 &{\footnotesize  }& {\footnotesize }& {\footnotesize }& {\footnotesize  -27.5 }& {\footnotesize 3.9 }& {\footnotesize -0.18  }& 237.898318 & {\footnotesize   -19.8 }& {\footnotesize 11.4 }& {\footnotesize 1.04 }& {\footnotesize  }& {\footnotesize }& {\footnotesize   }\\
\hline
\end{tabular}
\end{center}
\end{table*}

\setcounter{table}{0}

\begin{table*}
\caption[]{Continued.}
\begin{center}
\begin{tabular}{|lr|ccccccc|ccccccc|}
\hline
{\footnotesize V. state} & {\footnotesize type of }& \multicolumn{7}{c|}{{\footnotesize J=12-11} }& \multicolumn{7}{c|}{{\footnotesize J=26-25}} \\
{\footnotesize {\bf  [E cm$^{-1}$]} }& {\footnotesize  $\ell$ transition }& {\footnotesize  $\nu$ (GHz) }& {\footnotesize  v$_{1}$ }& {\footnotesize $\Delta$v$_{1}$ }& {\footnotesize A$_{1}$ }& {\footnotesize   v$_{2}$ }& {\footnotesize $\Delta$v$_{2}$ }& {\footnotesize A$_{2}$ }& {\footnotesize }  
{\footnotesize v$_{1}$ }& {\footnotesize  $\nu$ (GHz) }&{\footnotesize $\Delta$v$_{1}$ }& {\footnotesize A$_{1}$ }& {\footnotesize   v$_{2}$ }& {\footnotesize $\Delta$v$_{2}$ }& {\footnotesize A$_{2}$ }\\
\hline
{\footnotesize (0004) }& {\footnotesize {[ 0 0 0]E} }& 110.543693 & {\footnotesize  -21.3 }& {\footnotesize 10.9 }& {\footnotesize 0.2 }& {\footnotesize -27.4 }& {\footnotesize 5.2 }& {\footnotesize -0.35 }& 239.370230 & {\footnotesize -21.1 }& {\footnotesize 8.8 }& {\footnotesize 0.63 }& {\footnotesize  }& {\footnotesize }& {\footnotesize   }\\
{\footnotesize  [892] }& {\footnotesize {[ 0 0 2]F} }& 110.548972 & {\footnotesize  }& {\footnotesize }& {\footnotesize }& {\footnotesize }& {\footnotesize }& {\footnotesize }& 239.125296 & {\footnotesize }& {\footnotesize }& {\footnotesize }& {\footnotesize }& {\footnotesize           }& {\footnotesize        }\\
{\footnotesize }& {\footnotesize {[004]E+F}  }& 110.554367 & {\footnotesize  -21.3 }& {\footnotesize 7.1 }& {\footnotesize 8.9 }& {\footnotesize   -28.6 }& {\footnotesize 4.8 }& {\footnotesize -0.23 }& 239.510881 & {\footnotesize }& {\footnotesize }& {\footnotesize }& {\footnotesize }& {\footnotesize }& {\footnotesize                          }\\
{\footnotesize }& {\footnotesize {[ 0 0 2]E} }& 110.562578 & {\footnotesize  -19.9 }& {\footnotesize 10.5 }& {\footnotesize 0.6 }& {\footnotesize  }& {\footnotesize }& {\footnotesize }& 238.962431 & {\footnotesize }& {\footnotesize }& {\footnotesize }& {\footnotesize }& {\footnotesize }& {\footnotesize                          }\\
\hline
{\footnotesize (0012) }& {\footnotesize {[ 0 1 0]E} }& 109.989997 & {\footnotesize  }& {\footnotesize }& {\footnotesize }& {\footnotesize -27.4 }& {\footnotesize 5.2 }& {\footnotesize  -0.30 }& 238.254341 &{\footnotesize   -21.7  }& {\footnotesize 8.1 }& {\footnotesize  0.53 }& {\footnotesize  }& {\footnotesize }& {\footnotesize    }\\
{\footnotesize [944] }&{{\footnotesize [ 0-1 2]F} }& 110.035645 & {\footnotesize  }& {\footnotesize }& {\footnotesize }& {\footnotesize -28.2 }& {\footnotesize 4.5 }& {\footnotesize -0.18 }& 238.388217 &{\footnotesize    -20.2 }& {\footnotesize 9.3 }& {\footnotesize 0.64 }& {\footnotesize  }& {\footnotesize }& {\footnotesize    }\\
{\footnotesize }& {\footnotesize {[ 0 1 2]E+F} }& 110.097528 & {\footnotesize  -21.3 }& {\footnotesize 5.7 }& {\footnotesize 0.09 }& {\footnotesize  -27.8 }& {\footnotesize 5.6 }& {\footnotesize -0.34  }& 238526.950 &{\footnotesize  -22.5 }& {\footnotesize 8.5 }& {\footnotesize 0.58 }& {\footnotesize -31.2 }& {\footnotesize 6.1 }& {\footnotesize -0.28 }\\
{\footnotesize }& {\footnotesize {[ 0-1 2]E} }& 110.148765 & {\footnotesize  -21.3 }& {\footnotesize 5.0 }& {\footnotesize 0.06 }& {\footnotesize  -27.7 }& {\footnotesize 4.7 }& {\footnotesize -0.16      }& 238.631641 & {\footnotesize   -21.5 }& {\footnotesize 8.8 }& {\footnotesize 0.79 }& {\footnotesize }& {\footnotesize }& {\footnotesize  }\\
{\footnotesize }& {\footnotesize {[ 0 1 0]F} }& 110.189752 & {\footnotesize  }& {\footnotesize }& {\footnotesize }& {\footnotesize  -27.2 }& {\footnotesize 5.0 }& {\footnotesize -0.18   }& & {\footnotesize   -21.2 }& {\footnotesize 9.5 }& {\footnotesize 0.75 }& {\footnotesize  }& {\footnotesize }& {\footnotesize  }\\
\hline
{\footnotesize (0020) }& {\footnotesize {[ 0 0 0]E} }& 109.522442 & {\footnotesize  }& {\footnotesize }& {\footnotesize }& {\footnotesize -27.4 }& {\footnotesize 4.3 }& {\footnotesize -0.17 }& 237.270155 & {\footnotesize }& {\footnotesize }& {\footnotesize     }& {\footnotesize   }& {\footnotesize }& {\footnotesize    }\\
{\footnotesize [998] }& {\footnotesize {[ 0 2 0]E} }& 109.616120 & {\footnotesize  -21.3 }& {\footnotesize 7.5 }& {\footnotesize 0.23 }& {\footnotesize  -29.4 }& {\footnotesize 7.7 }& {\footnotesize -0.38     }& 237.468947 & {\footnotesize   -21.3 }& {\footnotesize 8.2 }& {\footnotesize 0.77 }& {\footnotesize  }& {\footnotesize e}& {\footnotesize    }\\
{\footnotesize } & {\footnotesize {[ 0 2 0]F} }& 109.616295 & {\footnotesize  -21.3 }& {\footnotesize 7.2 }& {\footnotesize 0.24 }& {\footnotesize  -28.8 }& {\footnotesize 8.3 }& {\footnotesize -0.41     }& 23.7470747 &{\footnotesize   -19.1 }& {\footnotesize 7.5 }& {\footnotesize 0.73 }& {\footnotesize  }& {\footnotesize }& {\footnotesize    }\\
\hline
{\footnotesize (1001) }& {\footnotesize {[ 0 0 1]E} }& 109.306687 & {\footnotesize  }& {\footnotesize }& {\footnotesize }& {\footnotesize  -27.0 }& {\footnotesize 6.0 }& {\footnotesize -0.17 }& 236.795297 &{\footnotesize  -20.0 }& {\footnotesize 4.5 }& {\footnotesize 0.10 }& {\footnotesize  -28.5 }& {\footnotesize 4.4 }& {\footnotesize -0.12 }\\
{\footnotesize [1103] }& {\footnotesize {[ 0 0 1]F} }& 109.469407 & {\footnotesize  }& {\footnotesize }& {\footnotesize }& {\footnotesize  -27.0 }& {\footnotesize 6.0 }& {\footnotesize -0.12 }& 237.145656 &{\footnotesize  -21.3 }& {\footnotesize 9.5 }& {\footnotesize 0.27 }& {\footnotesize  }& {\footnotesize }& {\footnotesize }\\
\hline
{\footnotesize (0102) }& {\footnotesize {[-1 0 2]E} }& 109.789853 & \multicolumn{6}{c|}{{\footnotesize blended with C$^{18}$O }     }& 237.803492 & {\footnotesize  -19.9 }& {\footnotesize 5.3 }& {\footnotesize 0.25 }& {\footnotesize }& {\footnotesize }& {\footnotesize      }\\
{\footnotesize [1109] }& {\footnotesize {[ 1 0 0]F} }& 109.847627 & {\footnotesize      }& {\footnotesize }& {\footnotesize }& {\footnotesize }& {\footnotesize }& {\footnotesize    }& 237.956672 & \multicolumn{6}{c|}{{\footnotesize  blended with (0002)[ 0 0 0]E}     }\\
{\footnotesize }& {\footnotesize {[ 1 0 2]E} }& 109.905021 & {\footnotesize      }& {\footnotesize }& {\footnotesize }& {\footnotesize }& {\footnotesize }& {\footnotesize    }& 238.164654 & {\footnotesize   -20.3 }& {\footnotesize 5.6 }& {\footnotesize 0.22 }& {\footnotesize }& {\footnotesize }& {\footnotesize      }\\
{\footnotesize }& {\footnotesize {[ 1 0 2]F} }& 109.868821 &{\footnotesize      }& {\footnotesize }& {\footnotesize }& {\footnotesize }& {\footnotesize }& {\footnotesize    }& 237.995859 & \multicolumn{6}{c|}{{\footnotesize  blended with (0002)[ 0 0 2]F}     }\\
{\footnotesize }& {\footnotesize {[ 1 0 0]E} }& 109.953634 &{\footnotesize      }& {\footnotesize }& {\footnotesize }& {\footnotesize }& {\footnotesize }& {\footnotesize   }& 238.178301 & {\footnotesize   -21.8 }& {\footnotesize 8.6 }& {\footnotesize 0.39 }& {\footnotesize }& {\footnotesize }& {\footnotesize     }\\
{\footnotesize }& {\footnotesize {[-1 0 2]F} }& 109.989048 & {\footnotesize      }& {\footnotesize }& {\footnotesize }& {\footnotesize }& {\footnotesize }& {\footnotesize   }& 238.318310 & {\footnotesize   }& {\footnotesize }& {\footnotesize }& {\footnotesize }& {\footnotesize }& {\footnotesize     }\\
\hline
{\footnotesize  (0005) }& {\footnotesize {[ 0 0 1]E} }& 110.655045 &{\footnotesize  -21.3 }& {\footnotesize 11.7 }& {\footnotesize 0.1 }& {\footnotesize -28.2 }& {\footnotesize 4.8 }& {\footnotesize -0.13 }& 230.471871 & {\footnotesize -21.3 }& {\footnotesize 7.9 }& {\footnotesize 0.23 }& {\footnotesize -28.5 }& {\footnotesize 4.6 }& {\footnotesize -0.14 }\\
{\footnotesize [1115] }& {\footnotesize {[ 0 0 1]F} }& 111.131162 &{\footnotesize  -21.3 }& {\footnotesize 15.6 }& {\footnotesize 0.26 }& {\footnotesize  -27.5 }& {\footnotesize 2.8 }& {\footnotesize -0.07 }& 231.456018 & {\footnotesize }& {\footnotesize }& {\footnotesize }& {\footnotesize }& {\footnotesize }& {\footnotesize                         }\\
{\footnotesize }& {\footnotesize {[ 0 0 5]E+F} }& 110.895474 & {\footnotesize        }& {\footnotesize      }& {\footnotesize      }& {\footnotesize  -28.2 }& {\footnotesize 4.8 }& {\footnotesize -0.13 }& 231.009985 & {\footnotesize }& {\footnotesize }& {\footnotesize }& {\footnotesize }& {\footnotesize }& {\footnotesize                         }\\
{\footnotesize }& {\footnotesize {[ 0 0 3]E+F}}& 110.902009 & {\footnotesize  -21.3 }& {\footnotesize  2.7 }& {\footnotesize  0.02 }& {\footnotesize   -28.2 }& {\footnotesize 5.0 }& {\footnotesize -0.13 }& 231.031080 &{\footnotesize }& {\footnotesize }& {\footnotesize }& {\footnotesize }& {\footnotesize }& {\footnotesize   }     \\
\hline
\end{tabular}
\end{center}
\end{table*}

\begin{table}
\caption[]{Set of models to describe the central continuum source 
of CRL618 according to the parameters defined in equation \ref{tcont}. 
The preferred is in bold face (see figure \ref{Fig1}).}
\begin{center}
\begin{tabular}{lrc}
size (``) & T$_{cont}$(200 GHz) & spectral index $s$ \\
\hline
0.18 &  8600 & -1.15 \\
0.22 &  6400 & -1.15 \\
{\bf 0.27} &  {\bf 3900} & {\bf -1.12} \\
0.32 &  3200 & -1.12 \\
0.36 &  2300 & -1.12 \\
\end{tabular}
\end{center}
\label{tablecont}
\end{table}

\begin{table}
\caption[]{Model parameters of expanding envelope that provide the 
best fit to the data shown in figure \ref{Fig6}. }
\begin{center}
\begin{tabular}{lrc}
parameter & value & units \\
\hline
$\theta_{g}$  & {\bf 1.5} & arcsec  \\
r$_{d}$  & {\bf 0.7} & -  \\
$i$ & {\bf 90} & degrees  \\
$T_{gas}$ (at $\theta$=$\theta_{c}$)  & {\bf 263} & K  \\
$T_{gas}$ (at $\theta$=$\theta_{g}$)  & {\bf 263}& K  \\
$v_{r}$ (at $\theta$=$\theta_{c}$)  & {\bf 5.0} & kms$^{-1}$   \\
$v_{r}$ (at $\theta$=$\theta_{g}$)  & {\bf 12.0} & kms$^{-1}$   \\
$v_{xy}$ (at $\theta$=$\theta_{c}$)  & {\bf 0.0} & kms$^{-1}$  \\
$v_{xy}$ (at $\theta$=$\theta_{g}$)  & {\bf 6.0} & kms$^{-1}$  \\
$v_{turb}$ (at $\theta$=$\theta_{c}$)  & {\bf 3.5} & kms$^{-1}$ \\
$v_{turb}$ (at $\theta$=$\theta_{g}$)  & {\bf 3.5} & kms$^{-1}$  \\
$[$HC$_{3}$N$]$ (at $\theta$=$\theta_{c}$)  & {\bf 154} & cm$^{-3}$   \\
$d$                                      & {\bf -1.8}  & -    \\
HC$_{3}$N col. density   & {\bf 2.0-3.5$\cdot$10$^{17}$} & cm$^{-2}$  \\
 at $p$$<$r$_{c}$ &  &  \\
\end{tabular}
\end{center}
\label{table2}
\end{table}

\clearpage

\end{document}